\documentclass[a4paper,11pt]{article}
\usepackage{pos}

\usepackage{listings}
\usepackage{xcolor}
\usepackage{standalone}
\usepackage{mathtools}
\usepackage{booktabs}
\usepackage[htt]{hyphenat} % automatic line breaks for textt env (\code{...} command)

\usepackage{tikz,ifthen}
\usetikzlibrary{calc}
\usetikzlibrary{math}
\usetikzlibrary{arrows.meta}
\usetikzlibrary{decorations.markings}
\usetikzlibrary{decorations}
\usetikzlibrary{positioning}

\definecolor{codegreen}{rgb}{0,0.6,0}
\definecolor{codegray}{rgb}{0.5,0.5,0.5}
\definecolor{codepurple}{rgb}{0.58,0,0.82}
\definecolor{backcolour}{rgb}{0.95,0.95,0.92}

\lstdefinestyle{mystyle}{
    backgroundcolor=\color{backcolour},   
    commentstyle=\color{codegreen},
    keywordstyle=\color{magenta},
    numberstyle=\tiny\color{codegray},
    stringstyle=\color{codepurple},
    basicstyle=\ttfamily\footnotesize,
    breakatwhitespace=false,         
    breaklines=true,                 
    captionpos=b,                    
    keepspaces=true,                 
    numbers=left,
    numbersep=5pt,                  
    showspaces=false,                
    showstringspaces=false,
    showtabs=false,                  
    tabsize=2
}

\definecolor{codegreen}{RGB}{0, 153, 0}
\definecolor{codegray}{RGB}{127, 127, 127}
\definecolor{codeblue}{RGB}{102, 214, 237}
\definecolor{codekeyword}{RGB}{249, 36, 114}
\definecolor{codecomment}{RGB}{127, 127, 127}
\definecolor{backcolor}{RGB}{242, 242, 235}
\definecolor{linkcolor}{RGB}{102, 0, 0}
\definecolor{corange}{RGB}{255, 70, 0}
\definecolor{cyellow}{RGB}{209, 153, 0}
\definecolor{cblue}{RGB}{64, 128, 255}
\definecolor{cbrown}{RGB}{153, 102, 51}
\definecolor{cpink}{RGB}{255, 0, 255}
\definecolor{cred}{RGB}{255, 64, 0}
\definecolor{cgreen}{RGB}{0, 191, 0}
\definecolor{clightblue}{RGB}{191, 217, 255}
\definecolor{cturquois}{RGB}{0, 255, 255}
\definecolor{cpurple}{RGB}{128, 0, 255}
\definecolor{clightgreen}{RGB}{175, 255, 175}
\definecolor{clightgray}{RGB}{211, 211, 211}
\definecolor{clightpink}{RGB}{255, 175, 255}
\definecolor{cdarkblue}{RGB}{0, 0, 255}
\definecolor{cdarkred}{RGB}{255, 0, 0}
\definecolor{cdarkgreen}{RGB}{0, 255, 0}

\def\code#1{\texttt{#1}}

\usepackage[noabbrev,capitalize]{cleveref}

\title{$O(a)$-improved QCD+QED Wilson Dirac operator on GPUs}
\author[a]{Anian Altherr}
\author[b]{Isabel Campos}
\author[c]{Jonathan Coles}
\author[d,e]{Alessandro Cotellucci}
\author[a]{Juan Antonio Fernández De la Garza}
\author*[a]{Roman Gruber}
\author[a]{Tim Harris}
\author[a]{Javad Komijani}
\author[d,f]{Jens Lücke}
\author[a]{Stephanie Maier}
\author[a]{Marina Krstić Marinković}
\author[a]{Letizia Parato}
\author[d,f]{Agostino Patella}
\author[b,d]{Sara Rosso}
\author[a]{Paola Tavella}
\author[c]{Hannes Vogt}

\affiliation[a]{ETH Zürich, Institute for Theoretical Physics, Wolfgang-Pauli-Str. 27, Zürich, Switzerlandntry}
\affiliation[b]{Instituto de Física de Cantabria and IFCA-CSIC, Avda. de Los Castros s/n, 39005 Santander, Spain}
\affiliation[c]{Swiss National Supercomputing Centre, CSCS, Lugano, Switzerland}
\affiliation[d]{Humboldt Universität zu Berlin, Institut für Physik and IRIS Adlershof, Zum Großen Windkanal 6, 12489 Berlin, Germany}
\affiliation[e]{Jülich Supercomputing Centre, Forschungszentrum Jülich, D-52428 Jülich, Germany}
\affiliation[f]{DESY, Platanenallee 6, D-15738 Zeuthen, Germany}
\affiliation[g]{Università di Roma Tor Vergata, Dip. di Fisica, Via della Ricerca Scientifica 1, 00133 Rome, Italy}
\affiliation[h]{INFN, Sezione di Tor Vergata, Via della Ricerca Scientifica 1, 00133 Rome, Italy}

\emailAdd{rgruber@ethz.ch}
\abstract{
Markov Chain Monte Carlo simulations of lattice Quantum Chromodynamics
(QCD) are the only known tool to investigate non-perturbatively the theory
of the strong interaction  and are required to perform precision tests of
the Standard Model of Particle Physics.
As the Markov Chain is a serial process, the sole option for improving the 
sampling rate is accelerating each individual update step.
Heterogeneous clusters of GPU-accelerated nodes offer large total memory
bandwidth which can be used to speed-up our application, openQxD-1.1,
which is dominated by inversions of the Dirac operator, a large sparse
matrix.
In this work we investigate offloading the inversion to GPU using the
lattice-QCD library QUDA, and our early results demonstrate a significant
potential speed-up in the time-to-solution for state-of-the-art problem
sizes.
Minimal extensions to the existing QUDA library are required for our
specific physics programme while greatly enhancing the performance
portability of our code and retaining the reliability and robustness of
existing applications in openQxD-1.1.
Our new interface will enable us to utilize pre-exascale infrastructure and
reduce the systematic uncertainty in our physics predictions by
incorporating the effects of quantum electromagnetism (QED) in our simulations.
}

\FullConference{The 41st International Symposium on Lattice Field Theory (LATTICE2024)\\
 28 July - 3 August 2024\\
Liverpool, UK\\}

\begin{document}
\maketitle

%--------------------------------------------------------------------------------
%--------------------------------------------------------------------------------
%--------------------------------------------------------------------------------
\section{Introduction}

%The goal of lattice Quantum Chromodynamics (lattice QCD) is to numerically simulate the interactions of quarks and gluons on a finite and discrete Euclidean spacetime grid, enabling first-principles predictions of hadron properties. 

%Computing hadronic observables to subpercent precision, as needed for example to interpret the most recent Muon g-2 measurements~\cite{Muong-2:2023cdq,Gerardin:2020gpp}, poses formidable challenges. It requires simulating four dynamical quark flavors, incorporating quantum electrodynamics (QED) effects, using fine lattice spacings for reasonable continuum extrapolation, using physical quark masses, and high statistics. Meeting these demands substantially increases computational requirements for generating Monte Carlo samples of gauge field configurations on large lattices.

The growing adoption of GPU-accelerated supercomputing in major facilities, largely driven by machine learning needs, presents new opportunities for using GPUs for lattice QCD codes. As lattice simulations are constrained by memory bandwidth rather than floating point operations, the high-memory throughput of GPUs is expected to result in a speed-up.

Our code, openQxD-1.1~\cite{openqxd}, extends the widely used openQCD-1.6 package~\cite{openqcd}, which supports the $O(a)$-improved Wilson fermion discretization (from now on, we will refer to these codebases simply as openQxD and openQCD, respectively). openQxD extends openQCD by allowing the simulation of QCD together with quantum electrodynamics (QED). This is accomplished through the use of C$^\star$ spatial boundary conditions \cite{Kronfeld1991}.
%, which respect locality and Gauss's law at the same time reducing finite-size effects compared to periodic boundaries or other lattice QED formulations.
%These features make openQxD a versatile tool for achieving higher levels of precision.
The codebase is designed for efficient parallel execution on CPU-based supercomputing architectures.

openQxD uses the Hybrid Monte Carlo algorithm for dynamical fermion lattice simulations, where the most time-consuming task is the repeated inversion of a large sparse operator to solve the Dirac equation. This Dirac operator, which connects nearest neighbor points on a 4D Euclidean spacetime lattice, has indices for Dirac spin, color charge, and space-time. By porting the Dirac solver to GPUs, the overall execution time of openQxD is expected to be significantly reduced.

Among the various possibilities for GPU-accelerating lattice QCD calculations, we choose to interface openQxD with the QUDA library~\cite{QUDApaper}, which provides optimized algorithms for NVIDIA and AMD GPUs, including an efficient iterative solver for the Dirac equation. 
We implement C$^\star$ boundary conditions and QED effects, as these are features that are particular to openQxD and do not yet exist in QUDA. In this way we combine openQxD's comprehensive physics toolset with QUDA's GPU-accelerated operations, thus enhancing openQxD simulations with GPU performance.

The remainder of this paper is structured as follows: \cref{sec:02-openqxd} introduces the openQxD codebase, while \cref{sec:03-quda} describes the QUDA library. 
\cref{sec:04-interface} details the process of interfacing the openQxD application with the QUDA library: it discusses the modifications to the memory layout of lattice fields, the implementation of $C^\star$ boundary conditions, and the extension of QUDA to handle QCD+QED simulations.
\cref{sec:07-performance} includes benchmark tests of CPU and GPU implementations of the solvers for the Dirac equation. Finally, \cref{sec:08-outlook} proposes future optimizations and summarizes the achieved GPU acceleration.

%--------------------------------------------------------------------------------
%--------------------------------------------------------------------------------
%--------------------------------------------------------------------------------
\section{openQCD}
\label{sec:02-openqxd}

\newcommand{\Dw}{{D_{\mathrm{w}}}}

The openQxD code~\cite{openqxd} is an extension of the state-of-the-art
lattice QCD Markov Chain Monte Carlo code
openQCD~\cite{openqcd,Luscher:2012av}, which is written in the C89 standard
with MPI parallelization with proven strong-scalability to hundreds of
CPU-based nodes and specific optimizations for x86 architectures.
Due to the extensive suite of legacy applications based on openQCD, as well as
its renowned stability and reproducibility, our goal is to accelerate the code
by offloading the solution of a large linear system of equations
$\Dw\psi=\eta$, defined by the Dirac operator $\Dw$, to the
device using the QUDA library.
The solution of this system for many right-hand sides is the most time
intensive kernel of the Monte Carlo algorithm, whose correctness can be
checked a posteriori, thereby allowing us to retain the functionality of
existing applications while interfacing only a minimal set of parameters.

The problem which we wish to offload is defined by the right-hand
side spinor $\eta$, which may be represented in different ways depending on 
conventions. These different conventions need to be mapped into each other in the interface.
Likewise, the couplings between the sites, which define the Dirac operator, are
parameterized by a set of so-called vector (or gauge) field variables $U_\mu$
for the four directions $\mu=0,1,2,3$, which need to be passed through the interface as well.
Although these fields are periodically updated along the Markov chain, in
the measurement part of the workflow they are kept fixed.
Finally, it is convenient to precompute the site-local coupling called the
clover-field which is a function of the gauge-field variables.
All these parameters are represented in openQCD by arrays of structs of length of
the local lattice volume.

The spinor field has indices $(x, \alpha, a)$ ($a$ running fastest), where $x=(x_0,x_1,x_2,x_3)$ is the lattice index (a Euclidean 4-vector), $\alpha \in \{0,1,2,3\}$ is the spinor index and $a \in \{0,1,2\}$ is the color index.

The gauge field instead has indices $(x_{odd}, \pm \mu, a, b)$, where $x_{odd}$ is the lattice index (only odd points, $V_\mathrm{L}/2$ elements), $\mu \in \{\pm 0, \pm 1, \pm 2, \pm 3\}$ is the direction of the gauge link (in positive as well as negative direction, i.e., 8 possibilities) and $a,b$ parameterize the row and column of the $SU(3)$-valued gauge link (row-major). Thus openQxD stores the gauge links in all $8$ directions at odd lattice points in contrast with most other common conventions which store $4$ gauge links in positive directions for every lattice point.

Finally, the clover field has indices $(x, \pm, i)$, where $x$ is the lattice index, $\pm \in \{+, -\}$ denotes the chirality (i.e. the upper ($+$) or lower ($-$) $6 \times 6$-block of the $12 \times 12$ clover matrix) and $i \in \{0, \dots, 35\}$ the $36$ non-zero real numbers needed to parametrize a $6 \times 6$ Hermitian matrix.

In addition to these data structs, already defined in openQCD, the openQxD package --
which provides an extension to incorporate electromagnetic gauge fields --
requires new functionality such as C$^\star$ boundary conditions
and additional degrees of freedom: these are described in more detail in
\cref{sec:qcd+qed,sub:cstar} and respectively.

%--------------------------------------------------------------------------------
%--------------------------------------------------------------------------------
%--------------------------------------------------------------------------------
\section{QUDA}
\label{sec:03-quda}

QUDA \cite{QUDApaper} is a library written in CUDA C++ that contains a suite of efficient kernels and solvers to implement lattice QCD simulations (the current release version 1.1.0 complies with the C++14 and 17 standards). It is thus natural that QCD fields are represented by objects, and the pointers to the fields' values are accessible via their objects.

A spinor field is represented by an instance of \code{ColorSpinorField}, a gauge field by \code{GaugeField} and a clover field as \code{CloverField}, all of which inherit from the \code{LatticeField} class. The different degrees of freedom are realised in the inheriting classes. The actual numbers are accessible via a pointer whose accessor is called \code{data()}.

QUDA is highly optimized to run on multiple GPUs, specifically on NVIDIA and AMD hardware. In addition, it offers a multi-threading backend.

%--------------------------------------------------------------------------------
%--------------------------------------------------------------------------------
%--------------------------------------------------------------------------------
\section{The interface}
\label{sec:04-interface}

Linking openQxD against QUDA involves three major steps:
1) The memory layout of lattice fields is a fundamental difference between the two applications. We implement an interface that reorders the fields to agree with the different conventions.
2) QUDA does not support $C^\star$ boundary conditions, so we implement them in the QUDA library.
3) QUDA does not support simulating QCD+QED either. This feature involves modifying the Wilson-Dirac operator and the clover term.

\subsection{Field reordering}

In order to use QUDA's functionality within openQxD, one has to build QUDA as a library, include the main header file (\code{quda.h}) and link openQxD against it. 
As an example for a QUDA function, we consider inversions of the Dirac operator, whose API call signature looks as follows:

\begin{lstlisting}[language=C++]
void invertQuda(void *h_x, void *h_b, QudaInvertParam *param);
\end{lstlisting}

Here, \code{h\_x} and \code{h\_b} are void pointers pointing to the host spinor fields in openQxD order, i.e. base pointers to arrays of \code{spinor\_dble} structs. The \code{QudaInvertParam} struct parametrizes the solver (see \code{quda.h} and Ref.~\cite{QUDApaper} for details).

We implement the reordering in the following classes in QUDA \cite{QUDApaper}:
\begin{itemize}
  \item \code{OpenQCDOrder} in \code{include/gauge\_field\_order.h} for the gauge field.
  \item \code{OpenQCDDiracOrder} in \code{include/color\_spinor\_field\_order.h} for the spinor field.
  \item \code{OpenQCDOrder} in \code{include/clover\_field\_order.h} for the clover field.
\end{itemize}

All of the fields have the Euclidean space-time index in common, so we begin by discussing the order of the lattice sites in the following section.

\subsubsection{Space-time coordinates}

Denoting the rank-local lattice extent in direction $\mu=0,1,2,3$ by $L_\mu \in \mathbb{N}$, we can write the lattice coordinate as a 4-vector, $x = (x_0,x_1,x_2,x_3)$, where $x_\mu \in \{ 0, \dots, L_\mu -1 \}$. openQxD puts time coordinate first $x = (t, \vec{x})$, which we refer to as (txyz)-convention. From that we can create a lexicographical index
\begin{equation} \label{eq:lexi}
\Lambda(x, L) := L_3 L_2 L_1 x_0 + L_3 L_2 x_1 + L_3 x_2 + x_3.
\end{equation}
openQxD orders the indices in so called cache-blocks; a decomposition of the rank-local lattice into equal blocks of extent $B_\mu \in \mathbb{N}$ in direction $\mu$. Within a block, points are indexed lexicographically $\Lambda(b, B)$ as in \cref{eq:lexi}, but the $L_\mu$ replaced by $B_\mu$ and $x$ replaced by the block local Euclidean index $b$, such that $b_\mu = x_\mu \mod B_\mu \in \{ 0, \dots, B_\mu -1 \}$.  
Furthermore, the blocks themselves are indexed lexicographically within the rank-local lattice decomposition into blocks, i.e. $\Lambda(n, N_B)$, where we denote the number of blocks in direction $\mu$ as $N_{B,\mu} = L_\mu / B_\mu$, and the Euclidean index of the block as $n$, such that $n_\mu = \lfloor x_\mu / B_\mu \rfloor \in \{ 0, \dots, N_{B,i} -1 \}$.

In addition, openQxD employs even-odd ordering, that is all even-parity lattice points (those where the sum $\sum_{\mu=0}^3 x_\mu$ of the rank-local coordinate $x$ is even) come first followed by all odd-parity points.

Therefore, the total rank-local unique lattice index is
\begin{align} \label{eq:ipt}
\hat{x} &= \biggl \lfloor \frac{1}{2} \Big( V_B \Lambda(n, N_B) + \Lambda(b, B) \Big) \biggr \rfloor + P(x) \frac{V}{2},
\end{align}
where $V_B = B_0 B_1 B_2 B_3$ is the volume of a block, $P(x)=\tfrac{1}{2}(1-(-1)^{\sum_\mu x_\mu})$ gives the parity of index $x$, and $b$, $n$ are related to $x$ as described in the text above.

This is implemented in openQxD by the mapping array \code{ipt}, $\hat{x} \coloneqq \text{ipt}\left[\Lambda(x,L)\right]$. Such mapping arrays are not recommended on GPUs due to shared memory contentions and memory usage. Instead, it is advisable to write a pure function $f \colon x \mapsto \hat{x}$ that implements \cref{eq:ipt} by calculating the index on the fly such that the compiler can properly inline the calculation.

We write the \code{OpenQCDDiracOrder} class that implements a \code{load()} and a \code{save()} method for the spinor fields. These two methods are called by QUDA in a loop with all possible values for \code{x\_cb} and \code{parity}, where \code{x\_cb} denotes the (local) checkerboard site index and \code{parity} the parity of the point. QUDA provides an (inlined) function \code{getCoords()} to translate the checkerboard and parity index into a (local) 4-vector $x = (\vec{x}, t)$ with time coordinate last ((xyzt)-convention). After permuting the coordinates into (txyz)-convention, we can query the mapping function $f$ to obtain for the desired lattice point the offset from the base pointer in openQxD and copy the data. That data might have additional indices that we describe in the
following.

\subsubsection{Spinor Field}

As mentioned previously, spinor fields have indices $(x,\alpha,a)$, where we describe how to transform the space-time index $x$ in the previous section. Both QUDA and openQxD use the same order for spinor index $\alpha$ and color index $a$. Thus at each space-time index we can copy $12=4 \cdot 3$ consecutive complex numbers (i.e. \code{24*sizeof(Float)} bytes where \code{Float} is a template parameter for real numbers (\code{double}, \code{float}, \code{half}) to or from the device. This is implemented in the order class \code{OpenQCDDiracOrder} in \code{include/color\_spinor\_field\_order.h} \cite{QUDApaper} and concludes spinor field reordering.

\subsubsection{Clover Field}

Similarly to the spinor field, for each space-time index, we can copy $72 = 2*36$ real numbers (i.e. \code{72*sizeof(Float)} bytes)
to the device (the save function is not necessary, since we never need to save clover fields from device to host).

openQxD stores the clover field as two arrays $u$ of length $36$ that represent two Hermitian 6$\times$ 6 matrices (one for each chirality $\pm$). The first $6$ entries are the diagonal real numbers and the following 15 pairs denote real and imaginary parts of the strictly upper triangular part in row-major order.

QUDA stores the strictly lower triangular part in column-major order. 
So we can transfer the clover field from openQxD to QUDA by specifying how these 36 numbers transform. In particular, the QED
clover field does not affect the block structure of the clover term and we can also transfer the clover term in QCD+QED (see \cref{sec:qcd+qed}
for details).
On the other hand, a pure QCD clover field can be calculated natively within QUDA and no additional transfer of fields is required in that case.

\subsubsection{Gauge Field}
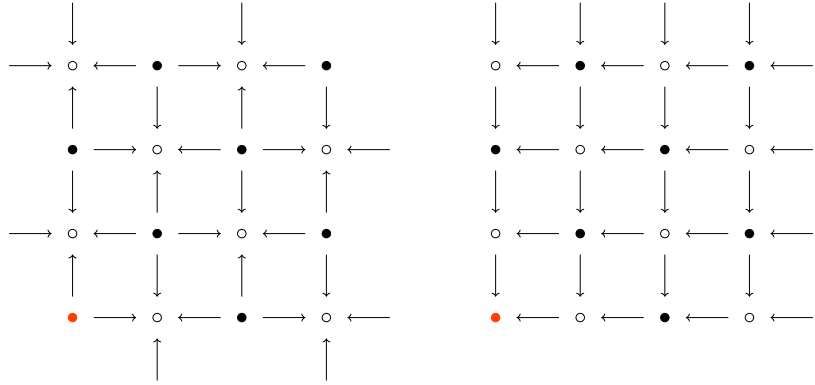
\begin{figure}
  \centering
  \begin{tikzpicture}

% openQxD
\foreach \x in {0,1,2,3} {
    \foreach \y in {0,1,2,3} {
        \coordinate (n) at ($\x*(20mm, 0)+\y*(0, 20mm)$);
        \coordinate (A1) at ($\x*(20mm, 0)+\y*(0, 20mm)+(0,15mm)$);
        \coordinate (B1) at ($\x*(20mm, 0)+\y*(0, 20mm)+(0,5mm)$);
        \coordinate (A2) at ($\x*(20mm, 0)+\y*(0, 20mm)+(15mm,0)$);
        \coordinate (B2) at ($\x*(20mm, 0)+\y*(0, 20mm)+(5mm,0)$);
        \coordinate (A3) at ($\x*(20mm, 0)+\y*(0, 20mm)-(0,15mm)$);
        \coordinate (B3) at ($\x*(20mm, 0)+\y*(0, 20mm)-(0,5mm)$);
        \coordinate (A4) at ($\x*(20mm, 0)+\y*(0, 20mm)-(15mm,0)$);
        \coordinate (B4) at ($\x*(20mm, 0)+\y*(0, 20mm)-(5mm,0)$);
        \ifthenelse{\x=0 \and \y=0} {
            \filldraw[color=cred,fill=cred](n) circle (0.1); % origin point
        } {
            \pgfmathparse{int(mod(\x+\y,2))};
            \ifnum\pgfmathresult=0 {
                \filldraw[color=black,fill=black](n) circle (0.1); % even points
            } \else {
                \draw[color=black,thin](n) circle (0.1); % odd points
                \draw[->] (A1) -- (B1);
                \draw[->] (A2) -- (B2);
                \draw[->] (A3) -- (B3);
                \draw[->] (A4) -- (B4);
            } \fi
        };
    };
};

% QUDA
\foreach \x in {0,1,2,3} {
    \foreach \y in {0,1,2,3} {
        \coordinate (n) at ($\x*(20mm, 0)+\y*(0, 20mm) + (100mm,0)$);
        \coordinate (A1) at ($\x*(20mm, 0)+\y*(0, 20mm)+(0,15mm) + (100mm,0)$);
        \coordinate (B1) at ($\x*(20mm, 0)+\y*(0, 20mm)+(0,5mm) + (100mm,0)$);
        \coordinate (A2) at ($\x*(20mm, 0)+\y*(0, 20mm)+(15mm,0) + (100mm,0)$);
        \coordinate (B2) at ($\x*(20mm, 0)+\y*(0, 20mm)+(5mm,0) + (100mm,0)$);
        \ifthenelse{\x=0 \and \y=0} {
            \filldraw[color=cred,fill=cred](n) circle (0.1); % origin point
        } {
            \pgfmathparse{int(mod(\x+\y,2))};
            \ifnum\pgfmathresult=0 {
                \filldraw[color=black,fill=black](n) circle (0.1); % even points
            } \else {
                \draw[color=black,thin](n) circle (0.1); % odd points
            } \fi
        };
        \draw[->] (A1) -- (B1);
        \draw[->] (A2) -- (B2);
    };
};
\end{tikzpicture}
  \caption{2D example ($4 \times 4$ local lattice) of how and which gauge fields are stored in memory in openQxD (left) and QUDA (right). Filled lattice points are even, unfilled odd lattice points. The red filled lattice point denotes the origin. Arrows represent gauge fields and the arrow head points to the lattice point where we associate the field with.}
  \label{fig:gauge}
\end{figure}

    QUDA associates 4 gauge fields for each space-time point (one for each positive direction $\mu=0,1,2,3$), whereas openQxD stores $8$
    (forward and backward) directions of gauge fields for all odd-parity points (see \code{main/README.global} for more information \cite{openqxd}). When looking at local lattices in a multi-rank scenario, this implies that openQxD locally stores gauge fields on the boundaries only for odd-parity points and not for even-parity points (see \cref{fig:gauge}). These even-parity boundary fields are stored in a buffer space, but they have to be communicated from neighbouring lattices first.

This requires us to transfer the missing gauge fields from one rank to the other before entering any QUDA interface function.

We implement an ordering class called \code{OpenQCDOrder} (see \code{include/gauge\_field\_order.h} \cite{QUDApaper}) with corresponding load and save methods. The only difference to the spinor and clover field order classes is that the methods are called with an additional direction variable \code{dir}. For a fixed space-time $x$ and direction, the remaining two color indices $(a,b)$ of the gauge field are row-major in both applications, thus we can copy $3\times 3$ consecutive complex numbers to or from the device.

\subsection{C\texorpdfstring{$^{\star}$}{*} boundary conditions}
\label{sub:cstar}

We choose to implement C$^\star$ boundary conditions in the same way as they are implemented in openQxD -- by doubling of the lattice in $x$-direction and by imposing shifted boundaries.

When QUDA initializes its communication grid topology, we specify the neighbours of each rank in all directions. The function \code{comm\_rank\_displaced()} in \code{include/communicator\_quda.h} \cite{QUDApaper} calculates the neighbouring rank number given one of (positive or negative) 8 directions. We change this function to achieve shifted boundary conditions

\subsection{QCD+QED}
\label{sec:qcd+qed}

In QCD+QED simulations, in addition to the $SU(3)$-valued gauge field $U_\mu(x)$, we have the $U(1)$-valued gauge field $A_\mu(x)$, which when combined results
in a $U(3)$-valued field $e^{i q A_\mu(x)} U_\mu(x)$ with $q_f$ the charge of a quark. These links are produced by multiplying the $U(1)$ phase to the $SU(3)$ matrices, which can be done in openQxD. For a QCD+QED operator in QUDA, we just upload these $U(3)$-valued links instead of the $SU(3)$ ones.

In addition, we add another SW-term proportional to its coefficient $c_{SW}^{U(1)}$. Therefore, the QCD+QED Clover term consists of both the $SU(3)$ and the $U(1)$ term, which can be calculated in openQxD. The resulting term is still diagonal in space-time and chirality, and is Hermitian in color and spin. Therefore it has the same representation in memory as the $SU(3)$ clover term alone and we can just upload this new clover field to QUDA using the clover field reordering class that we have already implemented.

Transferring the QCD+QED clover term, the $U(3)$ links as well as the changes in the process grid topology enables QUDA to handle the QCD+QED Wilson-Clover Dirac operator. This concludes our first implementation of QCD+QED in QUDA.

%--------------------------------------------------------------------------------
%--------------------------------------------------------------------------------
%--------------------------------------------------------------------------------
\section{Performance}
\label{sec:07-performance}

To obtain preliminary indications of the performance of the QUDA library on GPUs, we test on typical problem sizes and parameters relevant to our physics programme, see \cref{tab:ensembles}. For the comparison of the inversions between CPU and GPU implementations, we note that the multi-grid algorithm implemented in QUDA is not exactly the inexact deflation available in openQxD, therefore we anticipate further optimization of its parameters will improve its performance in future.

\begin{table}[h]
\centering
\begin{tabular}{ccccc}
Name & Global size $V_\mathrm{G}$ & Pion mass & Boundary conditions \\
\hline
G8   & $64^3 \times 128$ & $180$ MeV & periodic \\
C380 & $48^3 \times 96$  & $380$ MeV & C$^\star$ \\
\end{tabular}
\caption{The lattice ensembles used for the performance tests of the inversion of the Dirac operator. G8 has been generated by the CLS initiative~\cite{cls}, while C380 is generated by the RC$^\star$ collaboration~\cite{RCstar22}. The last column refers to spatial boundary conditions.}
\label{tab:ensembles}
\end{table}

\subsection{Solving the Dirac equation}

For solving the Dirac equation, we use a heavily preconditioned GCR algorithm. In case of the CPU runs, we use openQxD's deflated mixed-precision%
\footnote{single and double precision IEEE-754 floats} GCR solver that uses Schwarz alternating procedure as a preconditioner (\code{DFL\_SAP\_GCR} in openQCD parlance). The GPU runs use QUDAs mixed-precision%
\footnote{half, single and double precision IEEE-754 floats} multi-grid GCR solver (MG). The two solvers are very similar in how they generate the coarse grid subspace. The CPU solver parameters are tuned, but due to lack of experience with multi-grid, we used a somewhat not optimised parameter set for multi-grid on the GPU. The relative residual is chosen to be $10^{-12}$ for all solves.

\begin{figure}
    \centering
    \includegraphics[width=\linewidth]{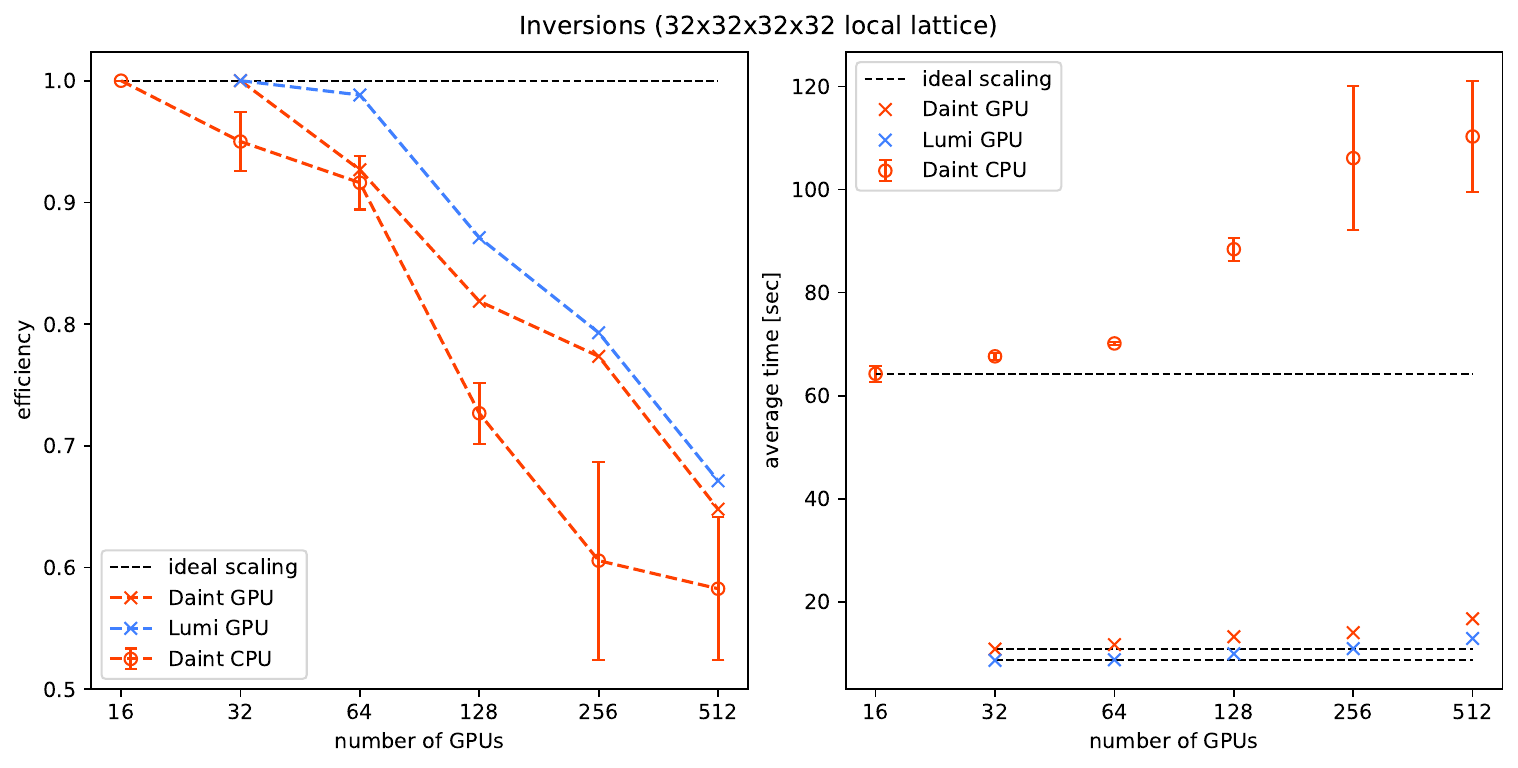}
    \caption{Weak scaling (left) and time to solution (right) for one solve of the Dirac equation on LUMI\protect\footnotemark (blue) and on Piz Daint (red) on the CPU\protect\footnotemark (circle) and GPU\protect\footnotemark (crosses) partition.}
    \label{fig:inv_weak}
\end{figure}

We obtain good weak scaling, see \cref{fig:inv_weak}. Again, LUMI outperforms Piz Daint in terms of cost. This is expected, since one LUMI-G node contains $4$ AMD MI250s (abstracted as $8$ devices) and one Piz Daint hybrid node contains a single NVIDIA P100. The factor is not 4x, but rather $\ge10$x, which is explained by the higher memory bandwidth of the former GPU\footnote{P100: 720 GB/s \cite{nv_p100} vs. MI250: 3.2768 TB/s \cite{amd_mi250}.}.
We note that strong-scaling for multi-grid algorithms is inherently challenging due to the reduced numbers of degrees of freedom on the coarser levels, which requires new algorithmic ideas~\cite{Espinoza-Valverde:2022pci}.

\footnotetext[4]{LUMI-C at CSC, Finland, 2x AMD EPYC 7763 @2.45GHz (2x 64 cores, 256-1024 GB RAM)}
\footnotetext[5]{Piz Daint multicore at CSCS, Switzerland, 2x Intel® Xeon® E5-2695 v4 @ 2.10GHz (2x 18 cores, 64/128 GB RAM)}
\footnotetext{Piz Daint hybrid at CSCS, Switzerland, 1x Intel® Xeon® E5-2690 v3 @ 2.60GHz (12 cores, 64GB RAM) and 1x NVIDIA® Tesla® P100 16GB}

%--------------------------------------------------------------------------------
%--------------------------------------------------------------------------------
%--------------------------------------------------------------------------------
\section{Outlook}
\label{sec:08-outlook}

In summary, we are able to use GPUs efficiently to solve the Dirac equation for a sample background gauge configuration from the Markov chain through our interface to the QUDA library.
The good performance observed with the QUDA multi-grid algorithm will allow us to perform efficient operations which underpin the measurement part of the workflow for fixed background gauge field and reduce the statistical and systematic precision on our lattice QCD predictions.
We plan to further develop and optimize this part (see  \cref{sec:optimizations}).

As a continuation of this project, we aim to port the generation of gauge configurations to GPUs. This will involve updating the gauge and clover field regularly, since these change periodically during the Markov chain. To solve this problem, and mitigate unnecessary transfer of fields from host to device, we design a powerful memory management system.

\subsection{Optimizations}
\label{sec:optimizations}

We plan to run simulations on Alps, the successor of Piz Daint at CSCS. This machine will have NVIDIA H100 GPUs. Unlike P100 GPUs, the H100 support NVSHMEM - a technology for inter-node communication in a multi-GPU scenario. Documentation claims excellent strong scaling of up to 1000 GPUs \cite{nvshmem}. QUDA supports this communication layer already and no new implementation effort has to be done to profit from it.

For the generation of gauge ensembles the gauge field changes frequently which will trigger recurring host-to-device (H2D) transfers of the field. To reduce CPU-GPU communication for QCD+QED simulations, we want to avoid a recomputation of the clover field on the CPU and a subsequent H2D transfer. We plan to implement the generation of the clover field on the GPU directly in QUDA.

In a second stage, we plan to unify the spinor fields among the two applications. The key idea is to not deal with fields on the CPU separately from fields on the GPU. A challenge is that the two applications have a vastly different memory layouts for fields and reordering has to be performed from one to the other. In openQxD a spinor field is just a large array of structs, whereas in QUDA a field is a C++ object. We plan to implement a memory management system, that keeps track the field status and implicitly transfer the field only if required.

%--------------------------------------------------------------------------------
%--------------------------------------------------------------------------------
%--------------------------------------------------------------------------------
\acknowledgments
We acknowledge access to Piz Daint at the Swiss National Supercomputing Centre, Switzerland under the ETHZ's share with the project IDs c21 and eth8 as well as LUMI at the CSC data center in Finland under the project IDs project\_465000469 and project\_465000803. We gratefully acknowledge PASC project "Efficient QCD+QED Simulations with openQ$^\star$D software". Finally, we thank Mathias Wagner and Kate Clark from NVIDIA as well as Jonathan Coles and Hannes Vogt from CSCS for the interesting discussions and their useful feedback on the development of this work.

%--------------------------------------------------------------------------------
%--------------------------------------------------------------------------------
%--------------------------------------------------------------------------------
\bibliography{include/references}

\end{document}